\documentclass[prl,aps,preprint,showpacs]{revtex4}
\usepackage{graphicx}

\begin{document}
  \title{Half-Metallic Ferromagnetism in  Transition-Metal Doped \\
Boron Nitride Nanotubes} 

  \author{H. J. Xiang}
  \author{Jinlong Yang }
  \thanks{Corresponding author. E-mail: jlyang@ustc.edu.cn}
  \author{J. G. Hou} 
  \author{Qingshi Zhu}

  \affiliation{Hefei National Laboratory for Physical Sciences at
    Microscale, Laboratory of Bond Selective Chemistry and Structure
    Research Laboratory, University of Science and Technology of
    China, Hefei, Anhui 230026, People's Republic of China}
  
  \date{\today}
  
  \begin{abstract}    
    We have studied zig-zag boron nitride (BN) nanotubes doped with
    the Ni hexagonal-close-packed nanowire. 
    The doped BN nanotubes are ferromagnetic metals with substantial
    magnetism.    
    Some special magnetic properties resulting from 
    the interaction between the Ni nanowire and BN nanotubes 
    are found. The Ni doped BN(9,0) nanotube
    shows semi-half-metallic behavior, which could become half-metallic
    after doping electrons more than 1.4 {\it e}/unit cell.
    The intrinsic half-metallic behavior could be
    achieved by two different ways: one is coating the Ni nanowire
    with a smaller BN 
    nanotube, i.e., BN(8,0), the other is using hydrostatic
    pressure to homogeneously compress the Ni doped BN(9,0) nanotube.
  \end{abstract}
  \pacs{75.75.+a,61.46.+w,72.25.-b,73.22.-f}
  \maketitle  
  Magnetic nanostructures are a scientifically interesting and technologically
  important area of research with many present and future applications in 
  permanent magnetism, magnetic recording and spintronics\cite{spintronics}. 
  One of the key properties in magneto electronics is the so-called half
  metallicity, i.e., conduction electrons that are $100\%$ spin polarized
  due to a gap at the Fermi level ($E_{f}$) in one spin channel, and a
  finite density of states (DOS) for the other spin channel. 
  Half metallic (HM) materials are ideal for spintronic applications,  such
  as tunneling  magnetoresistance  and giant
  magnetoresistance elements. 
  
  Encapsulating ferromagnetic structures inside nanotubes has become
  an active research topic.
  Carbon nanotubes
  filled with transition-metal(TM) have been studied extensively both
  experimentally
  \cite{cnt_mag_exp1,cnt_mag_exp2,cnt_mag_exp3,cnt_mag_exp4}
  and 
  theoretically\cite{cnt_mag1, cnt_mag2,cnt_mag3}. It was found that
  while TM nanowires filled carbon nanotubes can exhibit substantial magnetism,
  no HM behavior exists in such systems\cite{cnt_mag2}.

  Since boron nitride (BN) nanotubes are insulators independent of
  their chiralities and more chemically
  inert than carbon nanotubes, they may
  serve as naturally insulating and/or protective shields for
  encapsulating conducting metallic clusters, nanowires and
  nanorods\cite{BN_produce,BN_gap}. 
  The magnetic nanostructures isolated by a non-magnetic
  material, i.e., BN walls, could be efficiently used for high density
  data storage, without the drawback of particle agglomeration and
  magnetic loss due to dipolar relaxation.  
  This kind of protection also prevents the oxidation of the metallic
  clusters, nanowires and nanorods, which are prone to be oxidized.
  Recently,
  BN nanotubes filled with 3d-transition metals
  such as Fe-Ni Invar alloy, Co nanorods, Mo clusters, Ni and NiSi${_2}$
  nanowires have been
  successfully synthesized\cite{BN_TM1, BN_TM2,BN_TM3,BN_TM4}. 
  However, detailed electronic and magnetic properties of these
  novel nano-structures are far from well understood.
  And the question how the magnetic properties of TM/BN
  nanotubes differ from those of TM/C nanotubes is open.
  In this letter, we perform a comprehensive first
  principles study on the electronic and magnetic properties for 
  these TM/BN nanotube hybrid structures. 
  We show that the electronic
  and magnetic properties of TM/BN nanotubes differ fundamentally from
  those of TM/C nanotubes and    
  HM ferromagnetism could be achieved in TM/BN nanotubes under appropriate
  circumstances. 
  
  Our theoretical calculations are performed using the Vienna {\it
  ab initio} 
  simulation package (VASP)\cite{vasp1, vasp2}. We describe the
  interaction between ions and electrons
  using the  frozen-core projector augmented wave (PAW) approach
  \cite{paw, vasp_paw}. 
  The overall framework is spin-polarized density-functional theory (DFT)
  \cite{DFT1, DFT2} in the
  generalized gradient approximation (PW91)\cite{pw91_1, pw91_2}.
  The basis set cut off is $400$ eV.
  In a typical calculation,  a 1D periodic boundary condition is
  applied along the nanotube axis with Monkhorst-Pack\cite{mp} k-point
  sampling.  
  During structural relaxation,  all atoms except the fixed ones are
  allowed to relax to reach the minimum energies until the
  Hellmann-Feynman forces acting on them become less than $0.01$
  eV/\AA.
  Since most BN nanotubes are
  zig-zag type\cite{zig-zag1, zig-zag2},   
  we choose BN zig-zag nanotubes with a smallest hcp TM nanowire filled
  inside. The hcp TM nanowire is composed 
  of six TM atoms per unit cell with ABAB staggered triangle packing.
  BN(8,0), BN(9,0) and BN(10,0) nanotubes doped with the Ni hcp nanowire
  are studied.
  Different initial guesses are used for the local magnetic moments
  including ferromagnetic, antiferromagnetic (alternate up-down spins
  on A and B TM atoms), and nonmagnetic spin configurations, which
  are then fully relaxed to obtain the final converged structures and
  spin alignments. The free standing hcp Ni nanowire is also studied
  for comparison  with the hybrid structures. 
  
\begin{figure}
    \includegraphics[width=8.5cm]{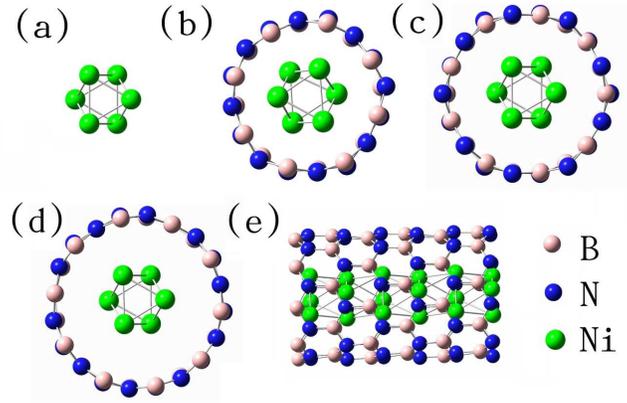}
    \caption{Relaxed structures of the free standing Ni nanowire and
      Ni/BN nanotube hybrid structures. (a) Top view of the free standing Ni
      nanowire. (b) Top view of Ni/BN(8,0). (c) Top view of
      Ni/BN(9,0). (d) Top view of Ni/BN(10,0). (e) Side view of
      Ni/BN(9,0). }
    \label{fig1}
  \end{figure}

  Fig.~\ref{fig1} shows the optimized structures for 
  the free standing Ni nanowire and 
  the Ni doped BN nanotubes. 
  In the optimized structures 
  Ni atoms almost lie in the nitrogen planes of BN nanotubes, as can be
  seen from the side view for Ni doped BN(9,0) nanotube
  (Fig.~\ref{fig1}(e)). This 
  relative position along axis between the BN nanotube and metal nanowire
  results from the big electronegativity of the nitrogen atom.
  The comparative calculation for Ni/C(9,0) shows that Ni atoms
  in Ni/C nanotubes don't lie in any carbon planes but in the middle of two
  adjacent carbon planes since all atoms are the same in carbon
  nanotubes. The symmetry of the Ni nanowire changes little after coated
  by BN nanotubes. Generally, after being inserted, the Ni nanowire expands a
  little, e.g., the Ni-Ni distance in a plane (A or B) changes
  from 2.27 \AA \ 
  to 2.32 \AA \ in Ni/BN(9,0), and the BN walls expand slightly, 0.01
  \AA\ for BN(9,0) nanotube.       
  
  As for the energetics of these hybrid structures,
  we give the formation energy of the hybrid structures in
  Table~\ref{table1}.   
  The formation energy is defined as
  $E_{b}=E(TM/tube)-E(TM)-E(tube)$.
  And the calculated formation energy for Ni/BN(8,0) is $0.88 $ eV, which
  implies the formation of this hybrid structure is endothermic. 
  The absolute value of the formation energy for Ni/BN(9,0) or
  Ni/BN(10,0) is very small. 
  The slightly favorable energy for Ni/BN(9,0) than
  that for Ni/BN(10,0) may result from the symmetry matching between the
  Ni nanowire and nanotube in Ni/BN(9,0).
  The
  positive or small negative formation energy is reasonable since the
  insulating BN 
  nanotubes are very stable and inert and there are significant
  difficulties in wetting a BN graphene-like surface.
  This also explains that the synthesized BN nanotubes filled with
  metals are limited and most of these hybrid structures are
  produced by a two-stage process\cite{BN_TM1, BN_TM2,BN_TM3}: first C
  nanotubes containing TM 
  nanoparticles at the tube-tips are synthesized, secondly,
  simultaneous filling nanotubes with the TM through
  capillarity and chemical modification of C tubular shells to form BN
  nanotubes occur. 
  But once the
  TM/BN nanotube hybrid structures are formed, they will be stable due to
  the large inertia of BN nanotubes.

  Since Ni/BN(9,0) is the most favorable form among the Ni doped BN
  nanotubes we studied, 
  we choose it as a typical case for studying the 
  detailed electronic and magnetic properties for Ni/BN
  nanotube structures. To serve as a reference, the electronic and magnetic 
  properties for the free standing Ni nanowire are also examined. 
  
\begin{figure}
    \includegraphics[width=8.5cm]{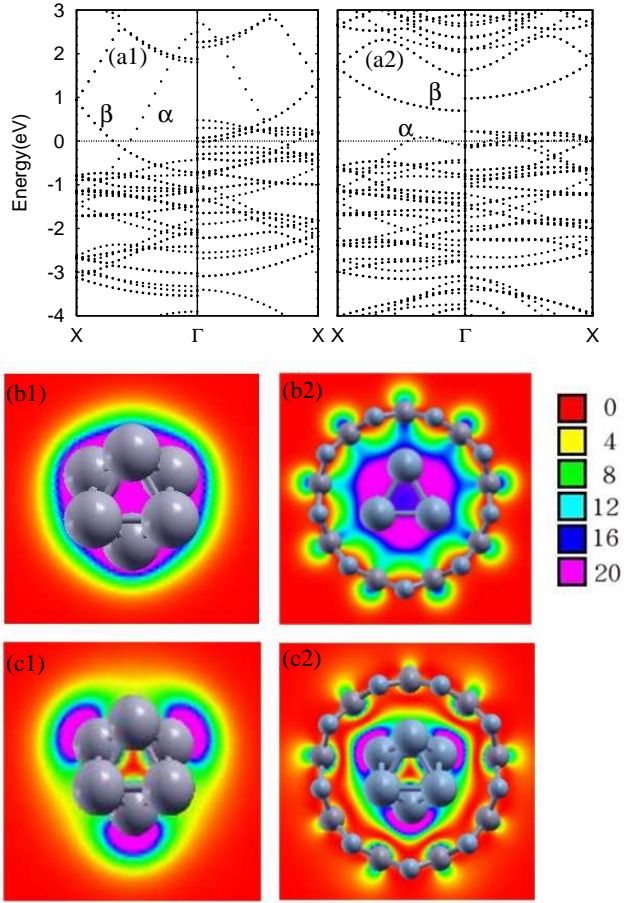}
    \caption{Spin resolved band structures and partial charges
      of the $\alpha$ and $\beta$ states for the free standing Ni nanowire and
      Ni/BN(9,0). (a1) and (a2) are the band structures for the free
      standing Ni 
      nanowire and Ni/BN(9,0).       
      The left panel is for the
      majority spin and the right panel is for the minority spin. 
      (b1) and (b2) are the partial charges
      for the $\alpha$ state of the free standing Ni nanowire and
      Ni/BN(9,0) respectively. (c1) and (c2) are the partial charges
      for the $\beta$ state of the free standing Ni
      nanowire and Ni/BN(9,0) respectively. 
      The $\alpha$ and$\beta$ states refer to those marked in (a1)
      and (a2). 
      }  
    \label{fig2}
  \end{figure}

  The band structures of  the free
  standing Ni nanowire and Ni/BN(9,0) are shown in Fig.~\ref{fig2}(a1)
  and (a2) respectively. 
  The free standing Ni nanowire is 
  ferromagnetic metal, but not HM or semi-HM,  as shown in
  Fig.~\ref{fig2}(a1). Similar results for the Co 
  nanowire were obtained by Yang {\it et al.}\cite{cnt_mag2}. 
  For the Ni nanowire and Ni/BN(9,0), the DOS around $E_{f}$ is always
  dominated by the minority spin part.
  The analysis of the projected density of states (PDOS) for Ni/BN(9,0)
  shows that the states around the Fermi level are mainly contributed by
  Ni 3d orbitals.
  And almost all the spin density 
  is located at the Ni nanowire. 
  So the electronic or
  spin transport will occur only in the core metal
  nanowires. 
  In the spin resolved band structure of Ni/BN(9,0), there is a gap
  just above the 
  Fermi level for the spin-up component indicating a semi-HM behavior,
  contrasting sharply with the  band structure for the free standing
  Ni nanowire. 
  It should be emphasized that the semi-HM
  behavior is found in the stablest ferromagnetic phase of TM/BN(9,0).
  Singh {\it et al.} also found the semi-HM
  behavior in a Mn doped Si hexagonal nanotube, however in the
  metastable ferromagnetic phase\cite{Si_TM}. The ferromagnetic ground
  state is helpful for application in spin-polarized transport.

  So why does the band structure for  Ni/BN(9,0) differ so much from that
  for  the free standing Ni nanowire since BN nanotubes are generally
  inert?    
  As the semi-HM energy gap arises in the spin majority part, we
  focus only on 
  the spin majority part. There are two states crossing the Fermi level
  in the spin majority part in the Ni nanowire, namely a non-degenerate
  state $\alpha$ and a double degenerate state $\beta$, as can be seen
  from  Fig.~\ref{fig2}(a1). Partial charge analysis shows that the $\alpha$
  state mainly distributes in the middle of planes A and B, while the 
  $\beta$ state almost locates at planes A and B.  
  After the Ni nanowire is inserted into BN nanotubes, the $\alpha$
  state near $\Gamma$ is
  shifted to lower energy, on the contrary, the $\beta$ state near
  $\Gamma$ is shifted to higher energy.
  To see what causes the change
  of the $\alpha$ and  $\beta$ states upon coated by BN nanotubes, we plot
  the partial charges of the $\alpha$ and $\beta$ states for the free
  standing Ni 
  nanowire and Ni/BN(9,0) in Fig.~\ref{fig2}(b) and (c). We
  can see 
  that the $\alpha$ state  becomes more delocalized but the $\beta$
  state  becomes more localized after the Ni nanowire is inserted into
  BN nanotubes. The different charge distribution between $\alpha$
  and $\beta$ states and the fact that the energy of $\alpha$
  state is lower than that of $\beta$ state in Ni/BN(9,0),
  suggest the $\alpha$ state is a bonding
  orbital and  the $\beta$ state is an anti-bonding orbital.
  The bonding and anti-bonding interactions are responsible for the
  different energy shifts for  the $\alpha$ and $\beta$ states
  respectively. 

\begin{figure}
    \includegraphics[width=8.5cm]{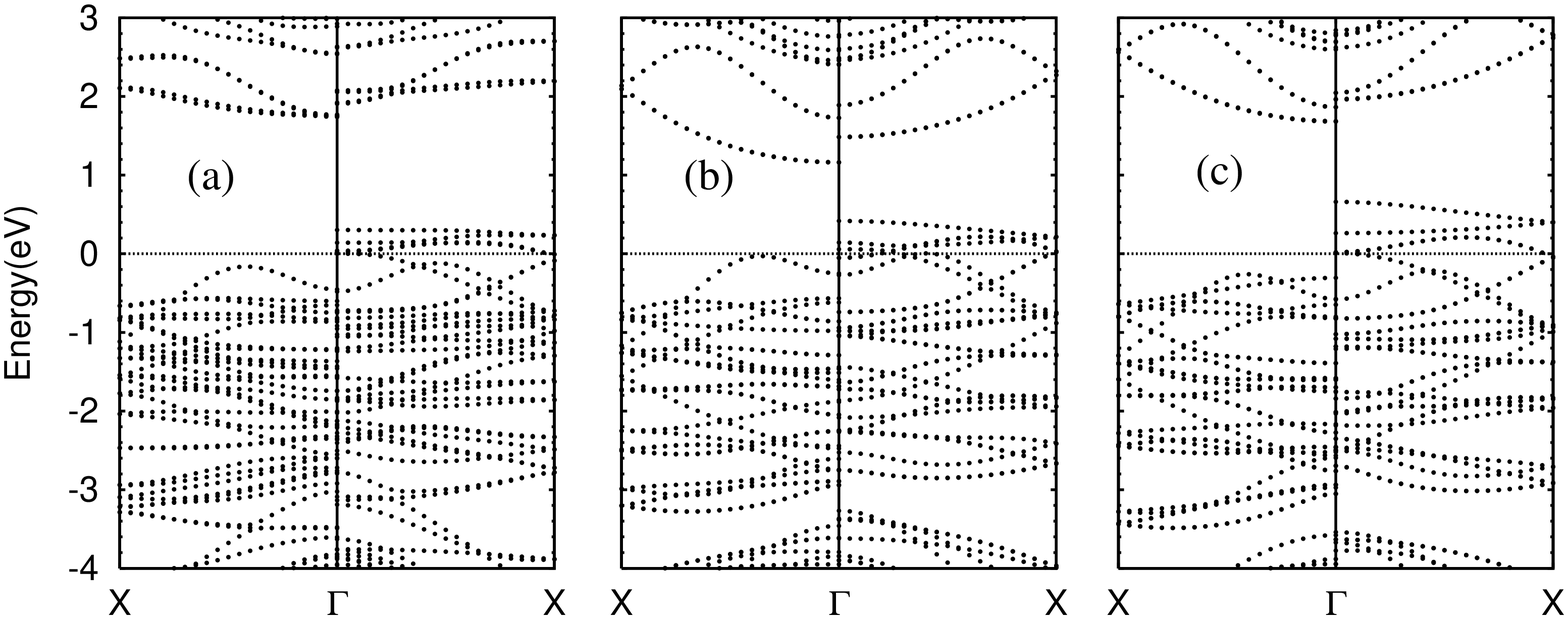}
    \caption{Spin resolved band structures for Ni/BN(8,0) and
      two radial homogeneously shrunken Ni doped BN(9,0) nanotube hybrid
      structures. (a) Ni/BN(8,0). (b) $5\%$ radial shrinkage. 
      (c) $10\%$  radial shrinkage.
      The left panel is for the
      majority spin and the right panel is for the minority spin. } 
    \label{fig3}
  \end{figure}

  Since low-dimensional HM materials are ideal for spintronic
  applications, can Ni 
  doped BN nanotubes become HM? 
  Theoretically, by doping electrons more than 1.4 {\it e}/unit cell
  to lift the Fermi level upward about 0.1 eV
  using some 
  techniques, such as applying gate voltage in a MOSFET like system,
  Ni/BN(9,0) could become HM.
  Even a metal to
  semiconductor transition could occur by doping 4.0 electrons/unit cell
  to lift the Fermi level upward more than 0.25 eV.
  On the other hand, seeking intrinsic HM in nano-structures is a more
  elegant solution to this question. 
  Since the semi-HM in Ni/BN(9,0) results from the hybridization
  between the Ni nanowire and BN nanotube, 
  the HM behavior is expected to occur in Ni doped BN nanotubes
  with an increased hybridization effect. 
  Two different means are considered to increase the hybridization.  
  First we study the hcp Ni nanowire coated with a smaller BN
  nanotube, i.e.,  BN(8,0) nanotube.
  The spin resolved band structure for Ni/BN(8,0)
  shown in  Fig.~\ref{fig3}(a) obviously indicates a HM behavior. 
  Though the formation of
  Ni/BN(8,0) is endothermic, Ni/BN(8,0) or other
  similar hybrid structures could be synthesized at high temperature
  with some subtle experimental methods, such as mechanic techniques
  and/or two-stage process.
  Secondly we investigate the Ni doped BN(9,0) nanotube 
  upon hydrostatic pressure. We simulate the pressure by
  fixing a homogeneous radial shrunken BN(9,0) nanotube and then
  fully relaxing the Ni nanowire. 
  Radial shrinkages to two different
  levels, $5\%$ and $10\%$, induced by homogeneous external pressures
  2.6 and 6.7 Gpa respectively, are examined. Here we assume that
  these external pressures would not cause Ni/BN(9,0) to transform to
  the oval shape. This is reasonable since the estimated transition
  pressure is 5.7 Gpa for BN(9,0) if we use the $1/R^3$ relationship
  between the transition pressure and the radius of the nanotube, and
  the fact that BN nanotubes have similar mechanical properties as carbon
  nanotubes, and the transition pressure for C(6,6) (about 4.0 Gpa)
  \cite{C_pres}, 
  moreover, the transition 
  pressure for Ni/BN(9,0) should be larger than this value due to the
  presence of the core Ni nanowire.   
  The band structures of the
  Ni doped shrunken BN(9,0) nanotubes are shown in Fig.~\ref{fig3}(b)
  and (c).
  We can clearly see that an energy gap is induced
  in the spin majority band of the hybrid structure after a
  small radial shrinking. Moreover, we find that the HM energy gap 
  increases along with the increase of the shrinkage, as seen from
  Fig.~\ref{fig3} that 
  the HM band gap for $10\%$ shrunken Ni/BN(9,0) is larger than that for
  $5\%$ shrunken Ni/BN(9,0). The radial shrinkage smaller
  than $5\%$ would not induce such a semi-HM to HM transition.
  In the two cases both with a decrease 
  in the radius of the nanotube and therefore increased
  hybridization, the bonding and 
  anti-bonding interactions become stronger. Once the interactions
  are strong enough, the HM can arise in these hybrid systems.
  Additionally, we find that the oval deformation couldn't induce the
  transition from semi-HM to HM.
  For the oval deformation, the hybridization between the Ni nanowire
  and BN nanotubes is not thorough since the oval deformation only
  increases the interaction between some Ni atoms and nanotube, but
  the interaction between other Ni atoms and nanotube is weakened. 
  
  The results about the total magnetic moment for the free standing Ni
  nanowire and hybrid Ni/BN nanotubes are listed in
  Table~\ref{table1}.  
  As we can see,
  the total magnetic moment are
  decreased after forming a hybrid Ni/BN nanotube. This is typically due to 
  the hybridization of the metal 3d states with the 2s and 2p states
  of boron and nitrogen.
  Interestingly, the reduction amount doesn't have a simple
  relationship with the radius of the BN nanotubes. When the radius
  is large enough, the reduction amount decreases with the increase of
  the nanotube radius, which can be explained by the weak hybridization
  effect. 
  
  Though, only Ni doped BN nanotubes are analyzed in detail in our
  discussion, the main results for Ni doped BN nanotubes apply equally
  to other TM, such as Fe and Co, doped BN nanotubes. All 
  these three TM/BN(8,0) are HM. The hydrostatic pressure can induce
  the transition from non-HM to HM for TM/BN nanotube hybrid
  systems. 
  By performing a calculation on the Ni nanowire coated by a
  double-walled BN nanotube, called BN(8,0)@BN(16,0), we find a HM
  behavior in such a system, 
  indicating that the speical magnetic properties for TM doped
  multi-walled BN nanotubes are mainly dependent on the innermost BN
  wall. Detailed discussions will appear elsewhere\cite{not_pub}.     
  It is noteworthy that 
  the special magnetic properties are unique for  TM/BN nanotube hybrid
  structures and are not found in previous studies on TM/C nanotube hybrid
  structures\cite{cnt_mag2} and the proposed systems consisting of B,
  N, and C atoms with a ferromagnetic ordering\cite{BCN1,BCN2}.

  In conclusion, we have performed an extensively first principles
  study on BN nanotubes doped with the Ni hcp nanowire. 
  All Ni/BN are
  ferromagnetic metals. The hybridization between the Ni nanowire
  and BN nanotubes leads to semi-HM or HM ferromagnetism in 
  Ni/BN nanotubes.  Ni/BN(9,0) is the most favorable form among the Ni doped
  BN nanotubes we studied and shows semi-HM behavior, which could be
  turned into HM behavior by doping electrons more than 1.4 
  {\it  e}/unit cell. 
  Moreover, HM ferromagnetism could be achieved by coating the Ni
  nanowire with a small radius BN nanotube, i.e., BN(8,0).
  Interestingly, for Ni doped BN nanotubes the
  HM behavior could also be 
  accomplished by the homogeneous shrinkage of BN nanotubes induced
  by hydrostatic pressure. And the HM band gap could be tuned by the
  applied pressure.
  Importantly, the interesting magnetic properties are not
  particular to Ni/BN nanotubes, but prevalent characters for Fe, Co
  and Ni doped BN nanotubes.    
  Almost all spin density and transport electrons locate in the core TM
  nanowire in TM/BN, differently from TM/C, in which electron and spin
  are more delocalized. All these special properties of TM/BN are very
  useful for spintronics and spin-polarized transport.

  This work is partially supported by the National Project for the
  Development of Key Fundamental Sciences in China (G1999075305,
  G2001CB3095), by the National Natural Science Foundation of China
  (50121202, 20025309, 10074058), by the Foundation of Ministry of
  Education of China, by the Foundation of the Chinese Academy of
  Science, and by the USTC-HP HPC project.

  \clearpage
  \begin{table} 
    \caption{The formation energy ($E_{b}$) and
      total magnetic moment ($\mu_{tot}$) for the free standing Ni
      nanowire 
      and several Ni doped BN nanotubes hybrid structures. Refer to
      the text for 
      the definitions for the formation energy ($E_{b}$).}
    \label{table1} 
    \begin{tabular}{c|c|c}
      \hline
      \hline
      &$E_{b}$(eV) &$\mu_{tot}$($\mu_{B}$/unit cell) \\ 
      \hline
      Ni nanowire&      &5.66 \\
      Ni/BN(8,0) &0.88  &4.00 \\
      Ni/BN(9,0) &-0.04 &3.49 \\
      Ni/BN(10,0)&0.03  &4.53 \\
      \hline
      \hline
    \end{tabular}
  \end{table}

  \clearpage

\end{document}